\documentclass[usenatbib]{mn2e}

\usepackage{amsmath}
\usepackage[dvips]{graphicx}

\title[New interpretation of expanding universe]{The New Interpretation of the Uniformly Expanding Universe}
\author[Y. Kubo]{Y. Kubo$^1$ \\
$^1$Japan Hydrographic Association, Tsukiji 5-3-1, Chuo-ku, Tokyo, 104-0045 Japan}

\begin{document}

\date{Accepted 2004 **** **. Received 2004 Aug **}

\maketitle

\begin{abstract}
The spacetime structure of the spatially uniformly expanding universe is described in terms of a kind of global space and
global time instead of the space and time we usually recognize.
The global space at some instant is a space in which the global time is equal and the global time is equal to
the proper time of any point moving together with the expansion of the universe which has elapsed since the big bang. 
The frame consisting of the global space and time thus defined is not rectangular when drawn in the usual spacetime framework.
In the new frame any expanding universe is open spatially and the Einstein's equations give a solution that
the space expands eternally regardless of the mass density of the universe indicating that any expanding universe is open dynamically as well,
as it is the case only when the density is less than the critical value in the standard model.
In fact the critical density has not any particular meaning in the new frame.
Finally it is shown that the equations related to the light path both in the standard model and in the new frame are similar.
\end{abstract}

\begin{keywords}
cosmology:theory --- large-scale structure of universe
\end{keywords}

\section{Introduction}

It is an observational fact described as the Hubble's law that the universe is expanding in a manner
that the speed of separation of any two points in the space is proportional to the distance between them.
We assume that the fact is true throughout the whole universe, that is, we accept the so-called cosmological principle.
This fact leads to a view that the space which we usually recognize as the space at the same instant as ours is not a space at a single instant
nor the time we regard as the same instant as ours belongs to this instant if the place is different,
suggesting the existence of a new frame which consists of a kind of global space and time coordinates.

The global space is so determined as to have the same value of time throughout that space, in which 
all the points are supposed to be comoving with the expansion of the universe.
On the other hand, the global time coordinate for any point which is moving with the expansion of the universe
is defined as the proper time at that point since the big bang
and thus the assembly of the points which have the same global time form a global space at the same instant.

The frame is essentially the same as one given by \citet[][p. 178]{ellis00}.
The spaces belonging to different times in this frame are expressed by various hyperbolas with the common asymptotes
and the time flow lines for all the points in the space are given by straight lines which start from the intersecting point of the asymptotes.
However, if we introduce an imaginary angle at the origin,
the spaces become concentric circles, providing us  with another new frame which is much simpler.

Rigorously the pictures of the both frames described above are valid only in the Milne universe or the massless universe
and a little different pictures must be drawn for the universe with mass.  Anyway, in the both types of the new frames 
the space and time coordinates system is not rectangular but similar to polar coordinates system
when referred to the framework of Minkowski world.

One of the conclusions with the new frame is that the volume of the space for any expanding universe 
is infinite or the universe is open statically
and also that the space in such a universe expands eternally with time
thus indicating that any expanding universe must be open dynamically as well.
This is a case only for a universe whose density is less than the critical density in the standard model.
But in the new frame this holds always regardless of the density of the mass and energy.

In Sections 2 to 3, the new framework of the universe which is essentially the same as that by Ellis \& Williams is introduced
in a different way from theirs, and in Section 4 it is shown how another type of new frame with an imaginary central angle can be introduced.
Also, in Section 5 it is shown that any universe is open spatially in the new frame whichever type we may adopt.

In Sections 6 and 7, the solution of the Einstein's equations is studied and compared with that in the standard model 
of Robertson-Walker spacetime with the result that the new solution always corresponds to the case
where the density is less than the critical density in the standard model.
This result comes from a fact that in the new frame the critical density is not critical
and no special phenomenon occurrs at this density.

Finally a further study on the nature of the new frame shows that the relation between
the space and time coordinates is quite similar to that in the ordinary spacetime
and thus the equations for the light path and the redshift, for example, are almost the same in the new and standard frameworks.

\section{Expanding Universe and Special Theory of Relativity}

Fig. 1 shows a Minkowski world of which the origin A is an observer or a galaxy $\textrm{G}_{\mathrm{A}}$ at some instant 
and $x^{0}$ axis or the time axis is $\mathrm{A}_{1}\mathrm{A}_{2}$, the world line of $\mathrm{G}_{\mathrm{A}}$.
We suppose that the galaxy is moving together with the average expansion of the universe without any deviation.
Note that the figure shows a Minkowski world with a one-dimensional space for simplicity.

The length of the segment for every time interval $\Delta t$ along the world line
of $\textrm{G}_{\mathrm{A}}$ as well as of any point in the space which is moving together with the expansion of the universe
is $ic\Delta t$, where $c$ is the speed of light.
We may express this fact as that the point proceeds along its world line with the speed $c$. 
 
Now let the temperature at A, which is represented by the cosmic background radiation there, be, say $2.7^{\circ}\mathrm{K}$.
Let $\mathrm{M}_{1}\mathrm{M}_{2}$ be the world line of a messenger who is dispatched from $\textrm{G}_{\mathrm{A}}$ 
and is at rest relatively to it at a distance of $d$ from it.
Consider now the case where the messenger observes the temperature at M, 
where the time is the same as at the origin A in the coordinates system at rest
 relatively to both the observer $\mathrm{G}_{\mathrm{A}}$ and the messenger.
The result of the observation is reported to $\mathrm{G}_{\mathrm{A}}$ and will be received after the time $d/c$.

The temperature which is observed by the messenger will be the same as that which is observed by an observer $\mathrm{G}_{\mathrm{B}}$ 
who is at M at the time of the observation but who is moving together with the expansion of the universe and 
therefore whose world line is $\mathrm{B}_{1}\mathrm{B}_{2}$.
We have $\angle \mathrm{M}_{2}\mathrm{BB}_{2} = \eta = \tan^{-1}(v/c)$, where $v$ 
is the speed of $\mathrm{G}_{\mathrm{B}}$ relative to $\mathrm{G}_{\mathrm{A}}$. 

The temperature thus observed and reported to $\mathrm{G}_{\mathrm{A}}$ by the messenger, however, 
will not be the same as that observed at A, i.e. $2.7^{\circ}\mathrm{K}$, but it will be, say $3.7^{\circ}\mathrm{K}$,
a little warmer than at A, 
for the reason as in the following:
$\mathrm{G}_{\mathrm{B}}$ will say that A is ahead of B concerning time and that $\mathrm{G}_{\mathrm{B}}$ did not observe the temperature 
at the same time as $\mathrm{G}_{\mathrm{A}}$ but at the instant when $\mathrm{G}_{\mathrm{A}}$ was at A$'$, earlier than A,
therefore it is quite natural that the temperature $\mathrm{G}_{\mathrm{B}}$ observed is warmer than at A.
Note that A$'$ is the event at the same instant as B for $\mathrm{G}_{\mathrm{B}}$ and $\angle$ABA$' = \eta$, too.

\begin{figure}
\includegraphics[width=84mm]{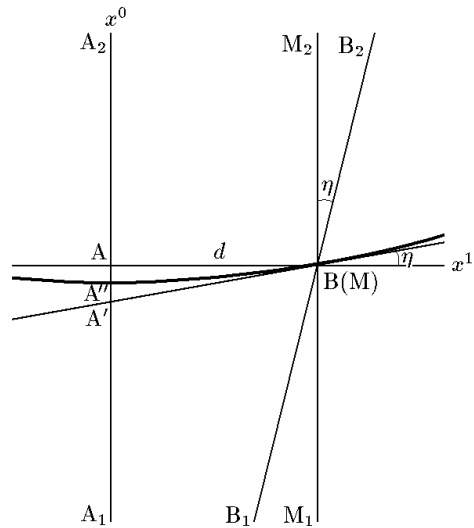}
\caption{Space with an equal temperature.  A$_{1}$A$_{2}$ and B$_{1}$B$_{2}$ are world lines of points
moving together with the expansion of the universe,
but M$_{1}$M$_{2}$ is that of a point at rest with respect to the point with the world line A$_{1}$A$_{2}$.
The temperature is the same on the curved line A$''$B.}
\end{figure}

However, if the roles of A and B are exchanged (in this case A$'$ but not A is to play the role of B),
$\mathrm{G}_{\mathrm{B}}$ insists that the temperature at A$'$ is the same as at B, $3.7^{\circ}\mathrm{K}$, 
while $\mathrm{G}_{\mathrm{A}}$ will say that the temperature at A$'$ is $4.7^{\circ}\mathrm{K}$.

This situation can be explained if we consider that the space with an equal temperature is curved to the direction of time 
in the ordinary spacetime which we recognize in such a way that the space with the same temperature as at A
is ahead referred to that for B by the amount of a half of AA$'$, as well as B is ahead by the same amount
referred to the space with the temperature at A$'$.
Thus the curve A$''$B is the space where the temperature is equal to $3.7^{\circ}\mathrm{K}$.

If we consider that the space with the same temperature has the same coordinate value of time,
it results that the space having the same value of time is curved in the spacetime framework which we usually recognize.
This picture results necessarily from the requirements of the special theory of relativity and of the fact that the universe is expanding,
or we can say that the special theory of relativity requires that the space of the expanding universe at some instant is curved
when expressed in the ordinary spacetime.
In conclusion we should regard the curve A$''$B as the true space of $\mathrm{G}_{\mathrm{B}}$
but A$'$B is only the space tangent to the true space at B.

From this view we should understand that the spacetime of the expanding universe has a structure
which can not be expressed by rectangular Minkowski world
but is represented better by a framework similar to polar coordinates system.
This is purely true for the expanding universe without mass.
The mass distribution may distort the frame considerably but still the frame would be closer to polar coordinates system
rather than to rectangular one.

\section{Global Time and Space at some Global Time}

We now introduce a global time which has a one-to-one correspondence with the temperature. 
In other words, we define such a global time as is equal throughout the space where the temperature is equal.
We will now consider the nature of the global time. 

In Fig. 2, let the extensions of the world lines of $\mathrm{G}_{\mathrm{A}}$ and $\mathrm{G}_{\mathrm{B}}$ intersect at O.
Then $\angle$AOB is $\eta, \eta$ being equal to $\angle \mathrm{M}_{2}\mathrm{BB}_{2}$ in Fig. 1.  Consider that $\eta$ is small enough
and that Fig. 2 is a Minkowski world with the origin at A.  AB$''$ is the space where the global time is equal to that at A.

\begin{figure}
\includegraphics[width=84mm]{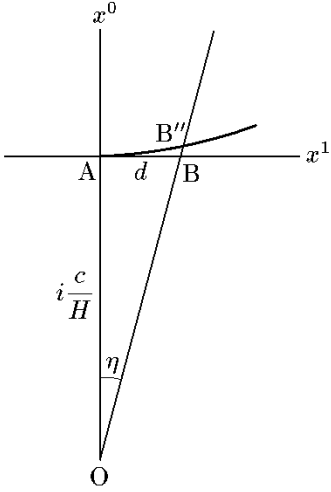}
\caption{Space at some global time.  The curve AB$''$ represents the space of the universe at some instant concerning the global time.
O is the centre for the curvature of the space.  In case of the Milne universe O corresponds to the big bang as well
and OA and OB$''$ correspond to the common proper time at A and B$''$ which has elapsed since the big bang.   }
\end{figure}

In the following, let $|\mathrm{OA}|, |\mathrm{AB}|$, etc. represent the Cartesian lengths of each segment
which would be obtained if the time axis were real.
Then they are all real.  On the other hand, denote their distances in Minkowski world by $\|\mathrm{OA}\|, \|\mathrm{AB}\|$, etc.

We have
\begin{equation}
\|\mathrm{OA}\| = i\frac{d}{\tan\eta} = i\frac{cd}{v} = i\frac{c}{H},
\end{equation}
where $H = v/d$ is the Hubble parameter at A and $d = \|\mathrm{AB}\|= |\mathrm{AB}|$.
If the universe is the Milne one where there is no mass, $v$ is constant and $d = vt$, $t$ meaning the time needed for A and B to separate up to the present distance.
This results $t = 1/H$, but in the non-Milne universe this is not true.

Since $\angle$B$''$AB $\simeq \eta/2$ in Fig. 2 after the discussion in the previous section, we have
\begin{equation}
|\mathrm{BB}''| \simeq \frac{1}{2}d\cdot\eta \simeq \frac{1}{2}\frac{c}{H}\eta^{2}. 
\end{equation}
Also we have
\begin{equation}
\|\mathrm{OB}\|^2 = \|\mathrm{OA}\|^2 + \|\mathrm{AB}\|^2 \simeq \frac{c^2}{H^2}(-1 + \eta^2).
\end{equation}
Hence,
\begin{equation}
\|\mathrm{OB}''\| = \|\mathrm{OB}\| + \|\mathrm{BB}''\| \simeq i\frac{c}{H}\sqrt{1 - \eta^2} + \frac{i}{2}\frac{c}{H}\eta^2
\simeq i\frac{c}{H}.
\end{equation}
Therefore we have $\|\mathrm{OB}''\| = \|\mathrm{OA}\|$ for $\eta$ small enough.
Note that $H$ is the Hubble parameter at B$''$ as well.

OA and OB$''$ are a kind of radii of the space AB$''$ and O is their centre.
In the Milne universe OA and OB$''$ are also the world lines of A and B$''$, respectively,
and O is the origin of the time coordinates, but generally this is not true.

Only in the massless universe the radii of the space and the world lines of the points in the space are coincident
and O is interpreted as the big bang.
And in this case $\|\mathrm{OA}\| = \|\mathrm{OB}''\| = ic\tau$, where $\tau$
is nothing other than the proper time which has elapsed since O.
Therefore we may define the global time $\hat{t}$ in the Milne universe by
\begin{equation}
\hat{t} = \tau = 1/H.
\end{equation}

As for the global time in the universe with mass, we can only say at this stage that the global time is equal if the value of $H$ is equal.
We will discuss about it later at the end of Section 6.  

Next we consider a hyperbola AC in Fig. 3 which is given by
\begin{equation}
(x^{0})^{2}+(x^{1})^{2}=\|\mathrm{OA}\|^{2}= -\frac{c^{2}}{H^2},
\end{equation}
$x^{0}$ being imaginary. $\mathrm{OZ}_{1}$ and $\mathrm{OZ}_{2}$ are the asymptotes of the hyperbola.  
Let C$(x^0, x^1)$ be an arbitrary point on the hyperbola.  Then
\begin{equation}
\|\mathrm{OC}\|^{2} = -\frac{c^{2}}{H^2} = \|\mathrm{OA}\|^{2}.
\end{equation}
Therefore the hyperbola AC is the space with an equal global time. 
If $\angle \mathrm{AOC} = \eta$ then $\tan \eta =v/c$, where $v$ is the speed of expansion at C relative to A.  
Here, $\eta$ is not necessarily small any longer.

\begin{figure}
\includegraphics[width=84mm]{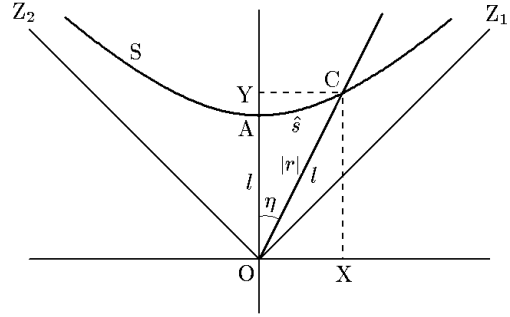}
\caption{Frame I, a frame to describe the spacetime structure of the expanding universe.
S represents the space at some instant and $l = ic/H$.  In the Milne universe $l = ic\hat{t}$ also holds 
and OA and OC are the world lines of points A and C, respectively, O corresponding to the big bang.
$|r|$ means the Cartesian length of OC.}
\end{figure}

Now we see that Fig. 3 gives a frame which describes the expanding universe correctly.
In the frame, hyperbolas having the same asymptotes as those for AC correspond to spaces at different times.
The straight lines passing O such as OA and OC are the radii of the space AC 
and in the massless universe they are the world lines of the points in the space as well.
We notice that this frame is essentially the same as one which is given by \citet{ellis00} for the Milne universe.

Next we will compute the length $\hat{s}$ of the curve AC in the Minkowski world.
Let the coordinates of C be $(x^{0},x^{1})$ and $|\mathrm{OC}|$ be $r$,
as well as $\|\mathrm{OA}\| = \|\mathrm{OC}\| = ic/H \equiv l$.  $r^{2}$ is obtained from 
\begin{equation}
(x^{0})^{2}+(x^{1})^{2}=-|\mathrm{CY}|^{2}+|\mathrm{CX}|^{2} = -r^{2}\cos^{2}\eta+r^{2}\sin^{2}\eta= l^2.
\end{equation}
Thus,
\begin{equation}
r = \frac{l/i}{\sqrt{\cos^{2}\eta - \sin^{2}\eta}}.
\end{equation}
From this,
\begin{equation}
x^{0} = \frac{l\cos\eta}{\sqrt{\cos^{2}\eta - \sin^{2}\eta}}, \ \ 
x^{1} = \frac{(l/i)\sin\eta}{\sqrt{\cos^{2}\eta - \sin^{2}\eta}},
\end{equation}
and
\begin{equation}
dx^{0} = \frac{l\sin\eta}{(\cos^{2}\eta-\sin^{2}\eta)^{3/2}}d\eta, \ \ 
dx^{1} = \frac{(l/i)\cos\eta}{(\cos^{2}\eta-\sin^{2}\eta)^{3/2}}d\eta.
\end{equation}
Then we have
\begin{equation}
d\hat{s} = \sqrt{(dx^{0})^{2}+(dx^{1})^{2}} = \frac{l/i}{\cos^{2}\eta - \sin^{2}\eta}d\eta,
\end{equation}
and thus
\begin{equation}
\hat{s} = \int^\eta_0\frac{l/i}{\cos^{2}\eta - \sin^{2}\eta}d\eta,
\end{equation}
$\hat{s}$ being real.

\section{Alternative Frame}

After \citet[][p. 191]{goldstein50} we now introduce an imaginary angle $\zeta$ defined by
\begin{equation}
\sin\zeta = \frac{\beta}{i\sqrt{1-\beta^{2}}}, \ \    \cos\zeta = \frac{1}{\sqrt{1-\beta^{2}}},
\end{equation}
where $\beta = \tan\eta = v/c$. Then
\begin{equation}
\tan\zeta = \frac{\beta}{i} = \frac{1}{i}\tan\eta,  
\end{equation}
and from this, 
\begin{equation}
d\zeta = \frac{1}{i}\frac{1}{1-\beta^{2}}\frac{d\eta}{\cos^{2}\eta} = \frac{1}{i}\frac{d\eta}{\cos^{2}\eta-\sin^{2}\eta}.
\end{equation}
Hence, from equations (13) and (16),
\begin{equation}
\hat{s} = l\int^{\zeta}_{0}d\zeta = l\zeta.
\end{equation}

Equations (7) and (17) strongly imply another and simpler frame to describe the spacetime structure of the expanding universe
as shown by Fig. 4.
We shall call the frame shown by Fig. 4 Frame II, while the frame given in Fig. 3 Frame I.

\begin{figure}
\includegraphics[width=84mm]{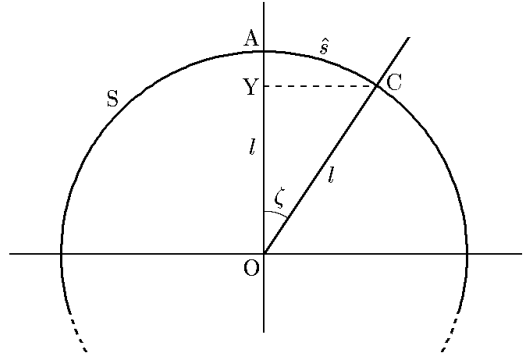}
\caption{Frame II, an alternative frame to describe the spacetime structure of the expanding universe.
S represents the space at some instant and it is the hyper-surface of the hyper-sphere.
OA and OC are the radii of the hyper-sphere and $l = ic/H$.
In the Milne universe $l = ic\hat{t}$ also holds and OA and OC are the world lines of points A and C, respectively, 
O corresponding to the big bang.}
\end{figure}

In the both frames it is quite natural to call $\hat{s}$ the global distance.
Meanwhile, $\eta$ or $\zeta$ is the global space coordinate making a counterpart of the global time $\hat{t}$,
proper to a point moving together with the expansion of the universe.
In Frame II the space at some instant $\hat{t}$ of the global time is a circle with the radius $l = ic/H$, 
and in the Milne universe the radii OA and OC are also the world lines of A and C, respectively, similar to in Frame I.

The global distance $\hat{s}$ between two points in the space, which is real,
is proportional to the angular distance $\zeta$.
While $\eta$ takes the values $-45^{\circ}$ to $+45^{\circ}, \zeta$ takes the values $-i\infty$ to $+i\infty$.
However, since the angle $\zeta$ is imaginary, neither $\sin\zeta$ nor $\cos\zeta$ has periodicity 
and therefore the circle does not close as in the case of real radius.

We will see the property of $\zeta$ more in detail.  We have from equation (14)   
\begin{equation}
e^{i\zeta}= \cos\zeta + i\sin\zeta = \sqrt{\frac{1+\beta}{1-\beta}},
\end{equation}
or 
\begin{equation}
\zeta = \frac{1}{2i}\left[\ln(1+\beta) - \ln(1-\beta)\right].
\end{equation}
Now, in the ordinary Minkowski world we consider a system $\mathrm{S}_{1}$ moving with the speed of $\varDelta u$ 
with respect to a system $\mathrm{S}_{0}$,
$\mathrm{S}_{2}$ moving with the speed of $\varDelta u$ in the same direction with respect to $\mathrm{S}_{1}$, and so forth.
Let the speed of $\mathrm{S}_{i}$ relative to $\mathrm{S}_{0}$ be $v_{i}$.
Then, from the formula for the addition of velocities in the special theory of relativity, we have   
\begin{equation}
\begin{split}
\varDelta v  &= v_{n} - v_{n-1} = \left(v_{n-1}+\varDelta u\right)\bigg/ \left(1+\frac{v_{n-1}\varDelta u}{c^{2}}\right)-v_{n-1}\\ 
          &= \left(1-\frac{v_{n-1}^{2}}{c^{2}}\right)\varDelta u \bigg/ \left(1+\frac{v_{n-1}\varDelta u}{c^{2}}\right).
\end{split}
\end{equation}
Dividing the both sides by $\varDelta u$ and replacing $\varDelta$ by $d$,
\begin{equation}
\frac{dv}{du} =1- \frac{v^{2}}{c^{2}}.
\end{equation}
From this,
\begin{equation}
\int^{u}_{0}{du} = c^{2}\int^{v}_{0}{\frac{dv}{c^{2}-v^{2}}}
= \frac{c}{2}\left[\ln\left(1+\frac{v}{c}\right) - \ln\left(1- \frac{v}{c}\right)\right].
\end{equation}
Comparing equations (19) and (22), we have
\begin{equation}
\zeta = \frac{1}{ic}\int^{u}_{0}{du}= \frac{u}{ic}.
\end{equation}

$u$ is interpreted as the velocity made by summing up all $\varDelta u$'s or integrating $du$ in the Galilean sense.
While $v$ can not exceed $c$, $u$ can be large to any amount.
Related to this, if we write the Hubble's law as $u = H\hat{s}$ instead of the usual expression $v = Hd$, then the law,
which holds only within a comparatively small distance in the current form, will hold completely throughout the universe.

\section{A Nature of the Space in the New Frame}

So far we considered only a one-dimensional space.  
So the space at some instant was a hyperbola in Frame I and a circle with imaginary radius and imaginary central angle in Frame II.
When we consider a two-dimensional space in Frame I, the space passing A comes to a hyperboloid 
which is made by rotating the hyperbola in Fig. 3 around the axis OA.
The set of the points with distance $\hat{s}$ from A make a circle with the radius $\|\mathrm{YC}\|$.

Similarly, if we proceed to the actual three-dimensional space, 
the space passing A comes to a hyher-hyperboloid and
the set of the points with distance $\hat{s}$ from A form a spherical surface with the radius $\|\mathrm{YC}\|$.

$\|\mathrm{YC}\|$ is given from equation (10) as
\begin{equation}
\|\mathrm{YC}\| = x^{1} = \frac{(l/i)\sin\eta}{\sqrt{\cos^{2}\eta-\sin^{2}\eta}}
 = \frac{(l/i)\beta}{\sqrt{1-\beta^{2}}} = l\sin\zeta.
\end{equation}
If $\zeta$ and $l$ are written as $i\bar{\zeta}$ and $i\bar{l}$, respectively, then, from equation (17),
\begin{equation}
\|\mathrm{YC}\| = i\bar{l}\cdot i\sinh\bar{\zeta} = \bar{l}\sinh \frac{\hat{s}}{\bar{l}}.
\end{equation}
Then the area $S$ of the spherical surface consisting of the points whose distance from A is $\hat{s}$ is given by
\begin{equation}
S = 4\pi\left(\bar{l}\sinh \frac{\hat{s}}{\bar{l}}\right)^{2},
\end{equation}
$S$ being real.
This is the very definition for that the  three-dimensional space is open.
  
In Frame II we can see equation (24) more directly.
In this frame, spaces of two-dimension and three-dimension are a spherical surface and the hyper-surface of a hyper-sphere 
with an imaginary radius, respectively.
$\|\mathrm{YC}\|$, which is real, is the radius of the circle for the two-dimensional space, and of the spherical surface 
for the three-dimensional space, each consisting of the points whose global distance from A is $\hat{s}$.

We can adopt the above discussion not only to the massless universe but also to the universe with mass as well.
Therefore it is concluded that any expanding universe is open spatially, that is, the volume of any such a universe is infinite.

\section{Einstein's Field Equations and the Evolution of the Radius \textit{\lowercase{l}}}

In the Milne universe, $l = ic/H$ evolves in such a way as $l = ic\hat{t}$.
However, when there is mass or energy, $l$ will not evolve like that
and in that case we have to solve Einstein's field equations to know how $l$ behaves with respect to time $\hat{t}$.

We start from the Einstein's field equations in Frame II.  We write the line element, which is given by
\begin{equation}
-c^{2}d\tau^{2} = -c^{2}d\hat{t}^{2} + l(\hat{t})^{2}(d\zeta^{2}+\sin^{2}\zeta d\theta^{2}+\sin^{2}\zeta\sin^{2}\theta d\phi^{2}),
\end{equation}
where $\tau, \hat{t}, \theta$ and $\phi$ are real but $l$ and $\zeta$ are imaginary.  
$l$ is considered to be a function of $\hat{t}$.
$l$ means the radius of hypersphere for the space at time $\hat{t}$, or $l = \hat{s}/\zeta$.  

On the other hand, the line element with Frame I is, as seen from Fig. 3,
\begin{equation}
\begin{split}
-c^{2}d\tau^{2} = &(dx^{0})^{2} +(dx^{1})^{2} +(dx^{2})^{2} +(dx^{3})^{2}\\ 
                = &[id(r\cos\eta)]^{2} +  [d(r\sin\eta\cos\theta)]^{2}\\
                  &+ [d(r\sin\eta\sin\theta\cos\phi)]^{2} \\
                  &+ [d(r\sin\eta\sin\theta\sin\phi)]^{2}. 
\end{split}
\end{equation}
After some calculations and putting $r = \bar{l}/\sqrt{\cos^{2}\eta - \sin^{2}\eta}$ due to equation (9), we have
\begin{equation}
\begin{split}
-c^{2}d\tau^{2} = &-c^{2}d\hat{t}^{2}\\
                  &+ \bar{l}(\hat{t})^{2}\left(\frac{d\eta^{2}}{\cos^{2}2\eta}
                   +\frac{\sin^{2}\eta}{\cos 2\eta}d\theta^{2}+\frac{\sin^{2}\eta\sin^{2}\theta}{\cos 2\eta} d\phi^{2}\right).
\end{split}
\end{equation}
However, due to equation (15), we see that equation (29) is equivalent to equation (27).

After the ordinary calculations, the Einstein's field equations for the line element written as equation (27) give
\begin{equation}
\frac{3}{l^{2}}(c^{2}+\dot{l}^{2})- \Lambda c^{2} = 8\pi G\rho,
\end{equation}
\begin{equation}
\dot{l}^{2}+2l\ddot{l}+c^{2} - \Lambda l^{2}c^{2} = -\frac{8\pi G}{c^{2}}l^{2}p,
\end{equation}
where $G$ is the universal gravitational constant, $\rho$ and $p$ are the density and the pressure, respectively, 
both considered constant spatially throughout the universe, and $\Lambda$ is the cosmological constant.
The dots over the variables represent the differentiation with respect to $\hat{t}$.

Now we limit our discussion to the case of $\Lambda = 0$ and $p = 0$.  Then we have from equations (30) and (31),
\begin{equation}
-\frac{2\ddot{\bar{l}}}{\bar{l}}= \frac{8\pi}{3}G\rho.
\end{equation}
Let
\begin{equation}
M = \frac{4\pi}{3}\bar{l}^{3}\rho.
\end{equation}
$M$ can not be regarded as the total mass of the universe, but we may suppose $M$ to be a constant.
Then from equations (32) and (33),
\begin{equation}
\frac{1}{2}\dot{\bar{l}}^{2} - \frac{GM}{\bar{l}} = E,
\end{equation}
$E$ being a constant.  But from equation (30),
\begin{equation}
\frac{1}{2}\dot{\bar{l}}^{2}-\frac{GM}{\bar{l}}=\frac{1}{2}c^{2},
\end{equation}
thus $E = c^{2}/2$.

We see from equation (35) that the universe continues to expand eternally once it expands, 
or the universe is open dynamically regardless of the mass density.

Solving equation (35) with the condition that $\bar{l} = 0$ at $\hat{t} = 0$, we have
\begin{equation}
c\hat{t} = \sqrt{\bar{l}(\bar{l}+2GM/c^{2})} + 
\frac{GM}{c^{2}}\ln{\frac{\sqrt{\bar{l}+2GM/c^{2}}-\sqrt{\bar{l}}}{\sqrt{\bar{l}+2GM/c^{2}}+\sqrt{\bar{l}}}}.
\end{equation}

Now we notice that the equation \begin{equation}
l = i \frac{c}{H}
\end{equation}
holds for any universe as we saw in Section 3.
Then from this equation and equation (33), we have
\begin{equation}
\frac{2GM}{c^2\bar{l}} = \frac{8\pi G}{3H^2}\rho = \frac{\rho}{\rho_{\textrm{cr}}} = \Omega,
\end{equation}
where $\displaystyle \rho_{\textrm{cr}} = \frac{3H^2}{8\pi G}$ is the so-called critical density, and therefore we have
\begin{equation}
c\hat{t} = \bar{l}\left[ \sqrt{1+\Omega}
+ \frac{\Omega}{2}\ln\frac{\sqrt{1+\Omega}-1}{\sqrt{1+\Omega}+1}\right].
\end{equation}
Though $\rho_{\textrm{cr}}$ is called the critical density in the standard model, $\Omega = 1$ is not any particular point in equation (39)
and $\rho_{\textrm{cr}}$ is not critical in the new frame.

By equations (37) and (38) and with the observed value of $H_{0}$ we can obtain $\bar{l}_{0}$ and $M$ 
for some value of $\Omega_0$, the subscript 0 signifying the value for the present time.
Then we can obtain the evolution of $\bar{l}$ with respect to $\hat{t}$ for various values of $\Omega_0$ from equation (39).

\begin{figure}
\includegraphics[width=84mm]{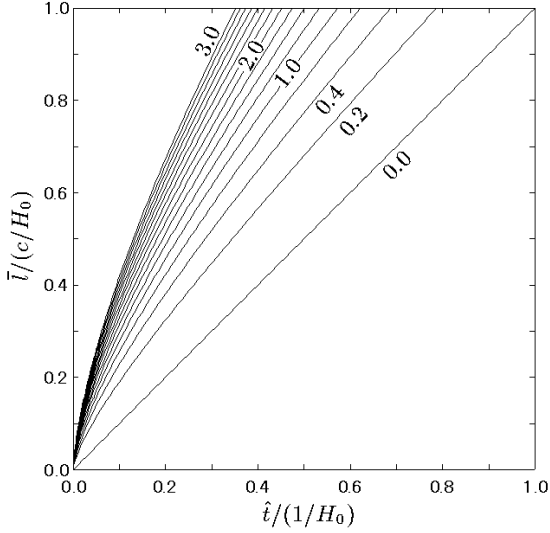}
\caption{Relation between $\hat{t}$ and $\bar{l}$.  The unit of $\hat{t}$ is $1/H_{0}$
and that of $\bar{l}$ is $c/H_0$, $H_{0}$ being the Hubble constant.
The numeral on each curve shows the value of $\Omega_0 = (\rho/\rho_{\mathrm{cr}})_0$.
The upper end of each curve corresponds to the present time.
$\Omega_0 = 1$ has not any special meaning.}
\end{figure}

The result is shown in Fig. 5. 
As inferred above, the critical density  has not any meaning in the new frame.  With  this density the universe ($\hat{s}$-$\hat{t}$ spacetime) is not flat
 and the age of the universe is about $0.53/H_0$, different from $2/(3H_0)$ which is the case in the standard model.
The age of the universe $2/(3H_0)$ occurs, on the other hand, when $\Omega_0 \simeq 0.46$, although this age is not especially significant.

Meanwhile, as for the evolution of the value $\Omega$ it is inversely proportional to $\bar{l}$ from equation (38), 
thus it was larger than unity in the past for any $\Omega_0$ other than 0.

Now we will discuss about the relation between $l$ and $\hat{t}$.
In Figs. 3 and 4, OA and OC are equal to $l$ in any universe but they are not equal to $ic\hat{t}$ except in the Milne universe.
Then where is the zero point of $\hat{t}$ or the big bang for the universe with mass in these figures? 
A possible interpretation on the difference between $l$ and $ic\hat{t}$ is given in Fig. 6.

\begin{figure}
\includegraphics[width=84mm]{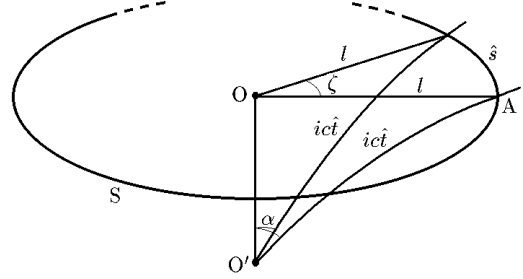}
\caption{A possible interpretation on the relation between $l$ and $\hat{t}$ in the universe with mass.  
OA is the radius of curvature of the space S and OA$ = l = i\bar{l}$ is imaginary. 
Meanwhile, O$'$A is the time flow for A and O$'$ corresponds to the big bang, $\alpha$ being $45^{\circ}$.
O$'$A $ = ic\hat{t}$ is also imaginary and, if OO$'$ is real, we have $\dot{\bar{l}} > c$ and so $\bar{l} > c\hat{t}$.}
\end{figure}

In the figure, S is the space at some instant.  Note that it is expressed in Frame II and as a one-dimensional space for simplicity.
O is the centre of the curvature of the space.
OA, the radius of curvature of S , is $l$ and it is imaginary.  On the other hand, the time flow of A is given by O$'$A,
O$'$ being the big bang.  $\alpha = 45^{\circ}$ and the curve O$'$A approaches to the horizontal direction asymptotically.  
Then O$'$A is imaginary and if OO$'$, which changes with time, is real we have $\dot{\bar{l}} > c$ and so $\bar{l} > c\hat{t}$.
A moves along the curve O$'$A with the speed $c$.  
The tangent to the curve O$'$A at A and OA are both perpendicular to the space S and that makes us feel they are the same.
So we will not be able to distinguish any universe from the Milne universe with the same value of $H$ from an instantaneous observation. 

\section{Comparison with Robertson-Walker Model}

We notice equation (27) is very similar to the line element in the Robertson-Walker spacetime 
which is regarded as the basis of the standard model of the universe.
The Robertson-Walker line element is written as
\begin{equation}
-c^{2}d\tau^{2} = -c^{2}dt^{2} + a(t)^{2}\left(\frac{dr^{2}}{1- kr^{2}}+r^{2}d\theta^{2}+r^{2}\sin^{2}\theta d\phi^{2}\right)
\end{equation}
\citep[][p. 74, e.g.]{peebles93}, where $a(t)$ is the expansion parameter which takes the value of unity for the present instant.

We may consider that equations (27) and (40) represent the same physical quantity expressed by different expressions.  In comparing them 
we first notice that there is no common variable to the both equations other than $\theta$ and $\phi$.
As for $t$ and $\hat{t}$, maybe they are the same at the observer's place or the origin of the space coordinate but not at other places.
And as for the other variables we have the following relations:
\begin{equation}
a(t)^{2}r^2 = l(\hat{t})^{2}\sin ^2 \zeta
\end{equation}
and
\begin{equation}
a(t)^{2}\frac{dr^2}{1- kr^2} = l(\hat{t})^{2}d\zeta^2.
\end{equation}
From these, if we assume that $k$ is a constant, we have
\begin{equation}
r = l_0\sin \zeta, \ \ a(t) = \frac{l(\hat{t})}{l(\hat{t}_0)} = \frac{l(\hat{t})}{l_0}, \ \ k = \frac{1}{l_0^2}. 
\end{equation}

Due to the last one of these equations, we can have $k = -1$ by choosing the unit of $l$ and $r$ appropriately,
but never $k = 0$ nor $k = +1$.
Only with $k = -1$, $a(t)$ behaves  in the same way as $l/l_0$.
This result is accordant with the conclusion which we have had above that $\bar{l} = l/i$ continues to increase eternally
regardless of the density of the universe.

Next we consider some problems related to the path of light in the new frame and compare them with those in the standard model.
Considering such a light as $d\theta = d\phi = 0$ in equation (27), we have
\begin{equation}
cd\hat{t} = ld\zeta.
\end{equation}  
First we examine the relation between the radius $l$ and the redshift $z$.
As for the light which was emitted from the point $\zeta$ at the time $\hat{t}_1$, when $l = l_1$, and is received at $\zeta = 0$
at $\hat{t}_0$, when $l = l_0$, we have from equation (44)
\begin{equation}
\zeta = -c\int_{\hat{t}_1}^{\hat{t}_0}\frac{d\hat{t}}{l}.
\end{equation}
Similarly, as for the light which was emitted from the same place as above at $\hat{t}_1+\Delta\hat{t}_1$ 
and is received at $\hat{t}_0+\Delta\hat{t}_0$, we have
\begin{equation}
\zeta = -c\int_{\hat{t}_1+\Delta\hat{t}_1}^{\hat{t}_0+\Delta\hat{t}_0}\frac{d\hat{t}}{l}.
\end{equation}
From these two equations we have
\begin{equation}
\frac{\Delta\hat{t}_1}{l_1} = \frac{\Delta\hat{t}_0}{l_0}. 
\end{equation}
Then 
\begin{equation}
1 + z = \frac{\lambda_0}{\lambda_1} = \frac{\Delta\hat{t}_0}{\Delta\hat{t}_1} = \frac{l_0}{l_1}.
\end{equation}
This is compared with the following equation in the standard model:
\begin{equation}
1 + z = \frac{\lambda_0}{\lambda_1} = \frac{a_0}{a_1}
\end{equation}
\citep[][p. 96, e.g.]{peebles93}. 
The similarity of equations (48) and (49) is a natural result from equations (43).

Next if we define $s$ by
\begin{equation}
s = a(t)\int _{0} ^{r} \frac{dr}{1 - kr^2},
\end{equation}
the behavior of $s$ with respect to $t$ is the same as that of $\hat{s}$ with respect to $\hat{t}$
except for the scale factors of $\hat{s}$ and $s$.
As a result, the expression for the path of light is the same both in $s$-$t$ frame and $\hat{s}$-$\hat{t}$ frame.
For example, in the latter frame, the equation for the path of the light which was emitted at $\hat{s}$ at the time $\hat{t}$
and reaches $\hat{s} = \zeta = 0$ at $\hat{t} = \hat{t}_0$ is written as 
\begin{equation}
\begin{split}
\hat{s} &= -l\zeta = l\int_{\zeta}^{0}d\zeta= cl\int _{\hat{t}}^{\hat{t}_0} \frac{d\hat{t}}{l}
         = c\bar{l}\int_{\bar{l}}^{\bar{l}_0}\frac{d\bar{l}}{\bar{l}\sqrt{c^2 + 2GM/\bar{l}}} \\
        &=\bar{l}\left[\cosh^{-1}\left(\frac{c^2}{GM}\bar{l} + 1\right)\right]_{\bar{l}}^{\bar{l}_0} \\
        &=\bar{l}\left[\cosh^{-1}\left(\frac{2}{\Omega} + 1\right)\right]_{\bar{l}}^{\bar{l}_0}.
\end{split}
\end{equation}
Also the horizon $\hat{s}_{\textrm{H}}$ at the present moment is given by
\begin{equation}
\hat{s}_{\textrm{H}_0} = \bar{l}_0\left[\cosh^{-1}\left(\frac{2}{\Omega}+1\right)\right]_0 ^{\bar{l}_0}
= \frac{c}{H_0}\cosh^{-1}\left(\frac{2}{\Omega_0} + 1\right).
\end{equation}

We have similar expressions for the standard model if we replace $\hat{s}$ and $\hat{t}$ by $s$ and $t$, respectively.
It should be noticed, however, that the quantity $s$ which corresponds to $\hat{s}$ in the standard model is not $a(t)r$
but is the quantity given by equation (50).  Besides, we know very little about $t$ for the universe with mass.
While we have a distinct definition of $\hat{s}$ and $\hat{t}$ as we saw in the present study,
we cannot say we know $s$ and $t$ with the same clarity.


\begin{thebibliography}{}
\bibitem [\protect\citeauthoryear{Ellis \& Williams}{2000}]{ellis00}%
  Ellis G. F. R. \& Williams R. M., 2000, Flat and Curved Spacetimes, 2nd edn. Oxford Univ. Press, Oxford
\bibitem [\protect\citeauthoryear{Goldstein}{1950}]{goldstein50}
  Goldstein H., 1950, Classical Mechanics. Addison-Wesley Publishing Company, Reading, MA
\bibitem [\protect\citeauthoryear{Peebles}{1993}]{peebles93}
  Peebles P. J. E., 1993, Principles of Physical Cosmology. Princeton Univ. Press, Princeton, NJ
\end{thebibliography}
\end{document}